\documentclass[preprint,prb,floatfix,showpacs,showkeys,superscriptaddress]{revtex4}

\usepackage{graphicx} %
\usepackage{subfigure} %
\usepackage{array} %
\usepackage{multirow} %
\usepackage{amsmath} %
\usepackage{float}
\usepackage{epstopdf}
\usepackage{dcolumn}
\usepackage[colorlinks=true,citecolor=blue,linkcolor=blue]{hyperref}%

\begin{document}

\title{First-principles study of ternary fcc solution phases from special quasirandom structures}

\author{Dongwon Shin}\email{dus136@psu.edu}
\affiliation{Department of Materials Science and Engineering, \\
The Pennsylvania State University, University Park, Pennsylvania 16802}
\author{Axel van de Walle}
\affiliation{Engineering and Applied Science Division, \\
California Institute of Technology, Pasadena, California 91125}
\author{Yi Wang}
\author{Zi-Kui Liu}
\affiliation{Department of Materials Science and Engineering, \\
The Pennsylvania State University, University Park, Pennsylvania 16802}

\date{\today}%

\begin{abstract}
In the present work, ternary Special Quasirandom Structures (SQSs) for a fcc
solid solution phase are generated at different compositions,
$x_A=x_B=x_C=\tfrac{1}{3}$ and $x_A=\tfrac{1}{2}$, $x_B=x_C=\tfrac{1}{4}$,
whose correlation functions are satisfactorily close to those of a random fcc
solution. The generated SQSs are used to calculate the mixing enthalpy of the
fcc phase in the Ca-Sr-Yb system. It is observed that first-principles
calculations of all the binary and ternary SQSs in the Ca-Sr-Yb system exhibit
very small local relaxation. It is concluded that the fcc ternary SQSs can
provide valuable information about the mixing behavior of the fcc ternary solid
solution phase. The SQSs presented in this work can be widely used to study the
behavior of ternary fcc solid solutions.
\end{abstract}



\maketitle

\section{Introduction}\label{sec:intro}

In CALPHAD (CALculation of PHAse Diagrams) modeling\cite{1970Kau,1998Sau},
thermodynamic properties of a solution phase in a ternary---or higher-order---
system are usually obtained through combining those of its constitutive binary
systems and occasionally with ternary interaction parameters. Since the most
dominant interatomic reaction in a multicomponent system is that of the
binaries, accurate thermodynamic descriptions which are capable of reproducing
the characteristics of binary solution phases are prerequisites to a successful
multicomponent thermodynamic modeling. In this regard, considerable efforts
have been made to develop models for combining thermodynamic descriptions for
binary solution phases to be used in higher-order
systems\cite{1960Koh,1965Too,1975Mug,1980Hil,1995Cho,2006Fan}.

In typical thermodynamic modelings, ternary interaction parameters for solid
solution phases are not introduced since their thermochemical data, such as
mixing enthalpy, which are needed in the parameter evaluation are very
difficult to measure. In fact, obtaining accurate thermochemical data for solid
solution phases is challenging even for binaries due to sluggish kinetics at
low temperatures so that it is hard to reach a complete thermodynamic
equilibrium. Furthermore, the existence of intermediate phases narrows the
experimentally accessible composition range for thermochemical properties of
solid solution phases. As the number of elements increases in a multicomponent
system, the complexity of acquiring reliable data also increases. Consequently,
interaction parameters for the excess Gibbs energy of binary solid solution
phases are usually evaluated only from phase equilibrium data and those of
ternary are usually omitted due to the lack of data.

Fortunately, the dearth of experimental data for solid solution phases can be
overcome by atomic level calculations, such as first-principles calculations.
Specially designed supercells, so-called Special Quasirandom Structures (SQS)
suggested by \citet{1990Zun}, whose lattice sites are occupied by constituent
atoms so as to reproduce the average correlation functions of a completely
random solution, are one of the representative methods. SQSs can be completely
relaxed within density functional theory (DFT) codes to consider the effect of
local relaxation and can also be applied to any other system by changing the
atoms because they are structural templates. A limitation is that SQSs of a
unit cell size that is manageable with present DFT codes can only be obtained
at certain compositions, e.g. $x$=0.25, 0.5, and 0.75 in the substitutional
$A_{1-x}B_x$ binary alloys. Nevertheless, first-principles study of SQSs at
those three compositions permit accurate predictions of various properties of
solid solutions. It has already been successfully applied to calculate
thermodynamic properties of binary solid solution phases for fcc, bcc and hcp
phases\cite{1990Wei,2004Jia,2006Shi}.

Two ternary fcc SQSs in an A-B-C system, which can be applied to four different
compositions at $x_A=x_B=x_C=\frac{1}{3}$; $x_A=\frac{1}{2}$,
$x_B=x_C=\frac{1}{4}$; $x_A=\frac{1}{4}$, $x_A=\frac{1}{2}$, $x_C=\frac{1}{4}$;
and $x_A=x_B=\frac{1}{4}$, $x_C=\frac{1}{2}$, are developed to investigate the
mixing enthalpies for ternary fcc solid solution phases in the present work.
The organization of this paper is as follows: the impact of ternary interaction
parameters on a ternary solution phase in the CALPHAD approach is briefly
reviewed. Then the generated ternary fcc SQSs are characterized in terms of
their atomic arrangements to reproduce the pair and multi-site correlation
functions of completely random fcc solid solutions. Finally, the generated SQSs
are applied to the Ca-Sr-Yb system which presumably has fcc solid solution
phases throughout the entire composition range in all three binaries and
ternary.

\section{Ternary interaction parameters}\label{sec:tern_interaction}
The Gibbs energy of a ternary solution phase, $\phi$, is expressed
as\cite{1980Hil}

\begin{equation}\label{eqn:ter_gibbs}
G^{\phi}=\sum^{c}_{i=1}x_iG^{o,\phi}_i + RT \sum^{c}_{i=1}x_i \ln x_i +
^{xs}G^{\rm bin,\phi}+^{xs}G^{\rm tern,\phi}
\end{equation}

\noindent where $x_i$ is the mole fraction of element $i$, $G^{o,\phi}_i$ is
the Gibbs energy of $\phi$ phase of pure element $i$, $^{xs}G^{\rm bin,\phi}$
and $^{xs}G^{\rm tern,\phi}$ are the excess Gibbs energies of the constitutive
binary and ternary systems, respectively. The excess Gibbs energies for binary
and ternary systems can be further described as

\begin{equation}\label{eqn:xs2_bin}
^{xs}G^{bin} = \sum^{c-1}_{i=1}\sum^{c}_{j>i}x_ix_j \sum^{n}_{\upsilon=0}
{}^{\upsilon} L^{\phi}_{ij}(x_i-x_j)^{\upsilon}
\end{equation}

\begin{equation}\label{eqn:xs2_tern}
^{xs}G^{tern} = \sum^{c-2}_{i=1}\sum^{c-1}_{j>i}\sum^{c}_{k>j}x_ix_jx_k (
L^{\phi}_ix_i+L^{\phi}_jx_j+L^{\phi}_kx_k)
\end{equation}

\noindent where ${}^{\upsilon} L^{\phi}_{ij}$ is the $v$-th order interaction
parameter\cite{1948Red} in a binary and normally described as

\begin{equation}\label{eqn:xs2_interaction}
{}^{\upsilon} L^{\phi}_{ij}= ^{\upsilon}a+^{\upsilon}bT
\end{equation}

\noindent where $^{\upsilon}a$ and $^{\upsilon}b$ are model parameters
evaluated from experimental information. Ternary parameter of element $i$,
$L^{\phi}_i$, in Eqn. \ref{eqn:xs2_tern} also have the form of Eqn.
\ref{eqn:xs2_interaction}.
If all three $L$-parameters in Eqn. \ref{eqn:xs2_tern} are identical, as in a
ternary regular solution\cite{1980Hil},

\begin{equation}\label{eqn:xs_identical}
L^{\phi}_{i}=L^{\phi}_{j}=L^{\phi}_{k}=L^{\phi}_{ijk}
\end{equation}

\noindent then the ternary excess Gibbs energy shown in Eqn. \ref{eqn:xs2_tern}
can be further simplified to

\begin{equation}\label{eqn:xs_simple}
^{xs}G^{\rm tern}=\sum^{c-2}_{i=1}\sum^{c-1}_{j>i}\sum^{c}_{k>j}x_ix_jx_kL^{\phi}_{ijk}
\end{equation}

\noindent since $x_i+x_j+x_k=1$ in a ternary.

\section{Ternary fcc special quasirandom structures}\label{sec:gensqs}
Thermodynamic properties of solid solution phases can be calculated in several
ways (e.g., see Refs. \cite{1991Duc,1994deF,2000Ced,2001Ast,2002van2,2002Col}).
A popular approach is to use a database of first-principles calculations to
determine the so-called Effective Cluster Interactions (ECI) that describe the
energetics of the alloy system of interest. These interactions are then used as
an input for efficient statistical mechanics methods, such as the Cluster
Variation Method or Monte Carlo simulations. While this general approach is
able to model ordered phases (with potential point defects) and disordered
(with potential short-range order) within a unified framework, it can be
computationally demanding. Fortunately, in cases where the sole objective is
obtain a reliable thermodynamic model for disordered phases that can be
reasonably assumed to lack significant short-range order, the use of SQS
provides a considerably more straightforward and computationally efficient
approach.

As discussed in Section \ref{sec:intro}, first-principles study of SQS can
effectively consider the limitations discussed above, and it has been
demonstrated that binary SQSs can be applied to calculate thermodynamic
properties of binary solid solutions, such as mixing enthalpy, for fcc, bcc,
and hcp structures\cite{1990Wei,2004Jia,2006Shi}. It is thus anticipated that
first-principles calculations of ternary SQSs should be able to reproduce
thermodynamic properties of ternary solid solutions if their atomic
configurations, which are represented as correlation functions, are very close
to those of ternary solid solutions. Correlation functions of solid solution
phases are well derived in \citet{1991Ind}. In the following section, the
correlation functions for binary and ternary systems are briefly summarized.

\subsection{Correlation functions}\label{sec:correlation_function}
The normalized correlation functions, $\overline{\Pi}_k$, in crystalline
structures are defined as

\begin{equation}\label{eqn:correlation_function}
\overline{\Pi}_k = \Pi^{c_1c_2 \dots c_k}_{12\dots k} =
\frac{1}{N}\sum_{k~site} \sigma^{c_1-1}_1 \sigma^{c_2-1}_2 \cdots
\sigma^{c_k-1}_k
\end{equation}

\noindent where the sum is over all the distinctive $k$-site clusters, which
are geometrically equivalent, in the $N$ lattice sites structure. When $k$=1,
2, 3, $\dots$, then $k$-site clusters are point, pair, triplets, and so forth.
Site occupation variables are denoted as $\sigma_k$, where the subscript $k$
indicates that the $k$-th constituent located in the corresponding site. The
superscript $c_k$ takes values 2, 3, $\cdots$, $C$, with $C$ as the number of
constituents, which represents a constituent $c_k$ at a given lattice site.

For binary systems when $C=2$, conventional values of the site occupation
variables $\sigma_k$ are $\pm$1 depending on whether a lattice site is occupied
by A or B atoms. According to Eqn. \ref{eqn:correlation_function}, the
normalized point correlation function for the second constituent site (B atom
sites) is given as $\Pi^2_1=\tfrac{1}{N_1}\sum\sigma^{2-1}_1$ with $\sigma_1$ =
1 or -1. It is worth noting that the atom sites do not need to be distinguished
in a binary system since they are switchable. With the mole fractions of A and
B being $x_A$ and $x_B$, respectively, ($x_A + x_B = 1$) then
$\Pi^2_1=x_A-x_B$. For a $k$-site cluster, the normalized correlation functions
for the binary solid solution is formulated as

\begin{equation}\label{eqn:cf_bin}
\overline{\Pi}_k=(x_A-x_B)^k
\end{equation}

Similarly, for ternary systems when $C=3$, the values of the site occupation
variables $\sigma_k$ conventionally take +1, 0, or -1 if a lattice site is
occupied by A, B, or C atoms, respectively. The normalized point correlation
function for the second constituent site (B atom sites) is given as
$\Pi^2_1=\tfrac{1}{N_1}\sum\sigma^{2-1}_1$ with $\sigma_1=$ +1, 0, or, -1. For
the third constituent site (C atom sites), the correlation function can be
given as $\Pi^3_1=\tfrac{1}{N_1}\sum\sigma^{3-1}_1$ with $\sigma_1=$ +1, 0, or,
-1. With the mole fractions of A, B, and C being $x_A$, $x_B$, and $x_C$,
respectively ($x_A+x_B+x_C=1$), then $\Pi^2_1=x_A-x_C$ and $\Pi^3_1=x_A+x_C$.
The vanishing of $x_B$ is due to its site occupation variable being 0. For a
$k$-site cluster with $n_B$ B atom sites and $n_C$ C atom sites ($n_B+ n_C=k$),
the normalized correlation functions for the ternary solid solution is denoted
as

\begin{equation}\label{eqn:cf_tern}
\overline{\Pi}_k=(x_A-x_C)^{n_B}(x_A+x_C)^{n_C}~~{\rm with}~~n_B+n_C=k
\end{equation}

\subsection{Generation of ternary fcc SQS}\label{subsec:gensqs}
In the present work, two different ternary fcc SQSs are generated. The first
SQS is at the equimolar composition where $x_A=x_B=x_C=\tfrac{1}{3}$ and the
second is at $x_A=\tfrac{1}{2}$, $x_B=x_C=\tfrac{1}{4}$. By switching the
occupation of the $A$ atoms in the second SQS with either $B$ or $C$ atoms, two
other SQSs can be obtained where $x_A=\tfrac{1}{4}$, $x_B=\tfrac{1}{2}$,
$x_C=\tfrac{1}{4}$ and $x_A=x_B=\tfrac{1}{4}$, $x_C=\tfrac{1}{2}$. Therefore,
mixing enthalpy at four different compositions in a ternary system can be
determined from first-principles total energy calculations of ternary fcc SQSs
by

\begin{equation}\label{eqn:enthalpy_of_mixing}
\Delta H(A_aB_bC_c)\approx E(A_aB_bC_c) - x_AE(A)-x_BE(B)-x_CE(C)
\end{equation}

\noindent where $E$ represents the total energy of each structure, and the
reference states for all pure elements are fcc.

When the number of atoms in the ternary SQS is less than 24, the Alloy
Theoretic Automated Toolkit (ATAT)\cite{2002van} has been used to generate
ternary fcc SQSs. Since the ATAT enumerates all the atomic configurations
within each supercell and then checks its correlation functions, the time
needed to find SQSs increases exponentially as the size of a supercell
increases. For the sake of efficiency, to find SQSs bigger than 24-lattice
sites, a Monte-Carlo-like scheme\cite{1997Abr} has been used. In each supercell
with different lattice vectors, atom positions are randomly exchanged between
the atoms and correlation functions of a supercell are calculated after every
alternation. If the correlation functions of a new state are getting closer to
those of random solutions, then the new configuration is accepted. Otherwise
the new state is discarded and another configuration will be generated from the
previous one. This process continues until the atomic arrangement of a
supercell converges to the closest correlation functions of the random
solution. In both methods, direct search via ATAT and Monte-Carlo-like scheme,
a supercell whose correlation functions matches best with that of a completely
random structure is chosen as the SQS at a given number of lattice sites.

\begin{figure}[htbp]
\centering %
    \subfigure[~SQS-24 when $x_A=x_B=x_C=\tfrac{1}{3}$]{%
        \label{fig:sqsABC}
        \includegraphics[height=2.25in]{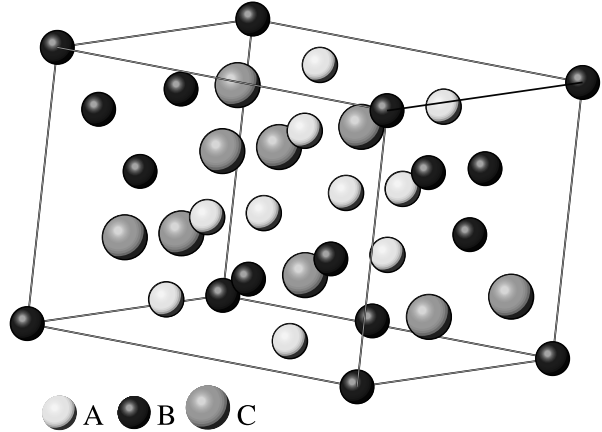}
    }
    \subfigure[~SQS-32 when $x_A=\tfrac{1}{2}$, $x_B=x_C=\tfrac{1}{4}$]{%
        \label{fig:sqsA2BC}
        \includegraphics[height=1.85in]{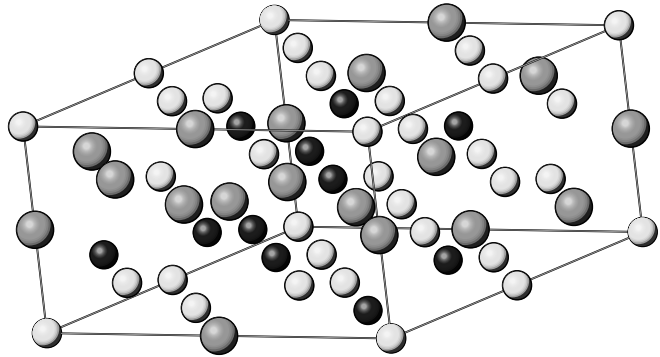}
    }
\caption{\label{fig:t_sqs}%
Atomic arrangements of the ternary fcc SQS structures in their ideal, unrelaxed
forms. All the atoms are at their ideal fcc sites, even though both structures
have the space group $P1$.}
\end{figure}

The selected SQSs at two different compositions, SQS-24 when
$x_A=x_B=x_C=\tfrac{1}{3}$ and SQS-32 when $x_A=\tfrac{1}{2}$ and
$x_B=x_C=\tfrac{1}{4}$, are shown in FIG. \ref{fig:t_sqs}. These two SQSs are
selected for later calculations because they are adequate with respect to their
size and correlation functions at each concentration\footnote{The criterion for
the total energy convergence with respect to different SQS sizes was set as 1
$m$ $eV/atom$.}. The space groups of both structures are $P1$ with all the
atoms at their ideal fcc sites. The correlation functions of the generated two
SQSs are given in TABLEs \ref{tbl:cfternABC} and \ref{tbl:cfternA2BC},
respectively.

\begin{table*}[htbp]
\centering%
\fontsize{7}{7pt}\selectfont %
\caption{\label{tbl:cfternABC}%
Pair and multi-site correlation functions of ternary fcc SQS structures when
$x_A=x_B=x_C=\tfrac{1}{3}$. The number in the square bracket next to
$\overline{\Pi}_{k,m}$ is the number of equivalent  at the same distance in the
structure, the so-called degeneracy factor.
}%
\begin{ruledtabular}
\begin{tabular}{lcdddddddd}
   & \multicolumn{9}{c}{SQS-$N$} \\
                           & Random & 3 & 6 & 9 & 15 & 18 & 24 & 36 & 48 \\
\hline %
$\overline{\Pi}_{2,1}$[6]  & 0 & 0 & 0 & 0 & 0 & 0 & 0 & 0 & 0 \\
$\overline{\Pi}_{2,1}$[12] & 0 & 0 & 0 & 0 & 0 & 0 & 0 & 0 & 0 \\
$\overline{\Pi}_{2,1}$[6]  & 0 & 0 & 0 & 0 & 0 & 0 & 0 & 0 & 0 \\
$\overline{\Pi}_{2,2}$[3]  & 0 & 0.25 & -0.125 & 0 & 0 & 0 & 0 & 0 & 0 \\
$\overline{\Pi}_{2,2}$[6]  & 0 & 0 & 0 & 0 & 0 & 0 & 0 & 0 & 0 \\
$\overline{\Pi}_{2,2}$[3]  & 0 & 0.25 & -0.125 & 0 & 0 & 0 & 0 & 0 & 0 \\
$\overline{\Pi}_{2,3}$[12] & 0 & -0.25 & -0.0625 & -0.0625 & 0 & -0.01042 & 0 & 0 & 0 \\
$\overline{\Pi}_{2,3}$[24] & 0 & 0 & 0 & 0 & 0 & 0 & 0 & 0 & 0 \\
$\overline{\Pi}_{2,3}$[12] & 0 & -0.25 & -0.0625 & -0.0625 & -0.06667 & -0.01042 & 0 & 0 & 0 \\
$\overline{\Pi}_{2,4}$[6]  & 0 & 0 & 0 & 0 & -0.075 & 0 & 0 & -0.04167 & 0\\
$\overline{\Pi}_{2,4}$[12] & 0 & 0 & 0 & 0 & -0.01443 & 0.0842 & 0 & 0 & -0.02255\\
$\overline{\Pi}_{2,4}$[6]  & 0 & 0 & 0 & 0 & -0.05833 & 0.09722 & 0.04167 & -0.04167 &  0.09896\\
\hline %
$\overline{\Pi}_{3,1}$[8]  & 0 & 0.125 & -0.01563 & 0.03125 & 0.04063 & 0.03125 & 0.01953 & -0.00391 & 0.01953\\
$\overline{\Pi}_{3,1}$[24] & 0 & 0 & 0 & 0 & -0.03789 & 0 & 0.01353 & 0.00226 & 0.00338\\
$\overline{\Pi}_{3,1}$[24] & 0 & -0.125 & 0.01563 & -0.03125 & 0.00938 & -0.03125 & -0.00391 & -0.02734 & 0\\
$\overline{\Pi}_{3,1}$[8]  & 0 & 0 & 0 & 0 & -0.00541 & 0 & 0.01353 & -0.00226 & 0.01691\\
$\overline{\Pi}_{3,2}$[12] & 0 & -0.125 & 0.0625 & 0 & -0.025 & 0 & -0.01562 & 0 & -0.00391\\
$\overline{\Pi}_{3,2}$[24] & 0 & 0 & 0 & 0 & 0.2165 & 0.01804 & 0 & 0.00902 & 0\\
$\overline{\Pi}_{3,2}$[12] & 0 & 0 & 0 & 0 & 0 & -0.03608 & 0.02706 & -0.01804 & 0.00677\\
$\overline{\Pi}_{3,2}$[24] & 0 & 0.125 & -0.0625 & 0 & -0.0125 & 0.01042 & 0.01563 & 0.01563 & -0.00781\\
$\overline{\Pi}_{3,2}$[12] & 0 & 0.125 & -0.0625 & 0 & -0.025 & -0.02083 & -0.01562 & -0.01042 & -0.02734\\
$\overline{\Pi}_{3,2}$[12] & 0 & 0 & 0 & 0 & 0 & 0 & 0.00902 & 0 & 0.00226\\
\end{tabular}
\end{ruledtabular}
\end{table*}

\begin{table*}
\fontsize{8}{8pt}\selectfont %
\centering%
\caption{\label{tbl:cfternA2BC}%
Pair and multi-site correlation functions of ternary fcc SQS structures when
$x_A=\tfrac{1}{2}$, $x_B=x_C=\tfrac{1}{4}$. The number in the square bracket
next to $\overline{\Pi}_{k,m}$ is the number of equivalent  at the same
distance in the structure, the so-called degeneracy factor.
} %
\begin{ruledtabular}
\begin{tabular}{lcddddddddd}
   & \multicolumn{8}{c}{SQS-$N$} \\
   & Random & 4 & 8 & 16 & 24 & 32 & 48 & 64 \\
\hline %
$\overline{\Pi}_{2,1}$[6]  & 0.0625 & 0.0625 & 0.0625 & 0.0625 & 0.0625 & 0.0625 & 0.0625 & 0.0625\\
$\overline{\Pi}_{2,1}$[12] & 0      & 0 & 0 & 0 & 0 & 0 & 0 & 0\\
$\overline{\Pi}_{2,1}$[6]  & 0      & -0.0625 & 0 & 0 & 0 & 0 & 0 & 0\\
$\overline{\Pi}_{2,2}$[3]  & 0.0625 & -0.125 & 0.0625 & 0.0625 & -0.0625 & 0.0625 & 0.0625 & 0.0625\\
$\overline{\Pi}_{2,2}$[6]  & 0      & 0 & 0 & 0 & 0.09021 & 0 & 0 & 0\\
$\overline{\Pi}_{2,2}$[3]  & 0      & 0.125 & 0.0625 & -0.0625 & 0.08333 & 0 & 0 & 0\\
$\overline{\Pi}_{2,3}$[12] & 0.0625 & 0.0625 & -0.00781 & 0.074219 & 0.0625 & 0.0625 & 0.0625 & 0.05957\\
$\overline{\Pi}_{2,3}$[24] & 0      & 0 & 0 & 0.006766 & 0.02255 & 0 & -0.002255 & 0.003383\\
$\overline{\Pi}_{2,3}$[12] & 0      & -0.0625 & -0.10156 & 0.011719 & 0.02604 & 0 & -0.002604 & -0.006836\\
$\overline{\Pi}_{2,4}$[6]  & 0.0625 & -0.125 & 0.015625 & 0.085938 & 0.15625 & 0.0625 & 0.0625 & 0.361328\\
$\overline{\Pi}_{2,4}$[12] & 0      & 0 & 0 & -0.040595 & 0 & -0.05413 & 0.009021 & -0.010149\\
$\overline{\Pi}_{2,4}$[6]  & 0      & 0.125 & -0.046875 & -0.023437 & 0.07292 & 0 & -0.03125 & 0.193359\\
\hline %
$\overline{\Pi}_{3,1}$[8]  & -0.015625 & -0.015625 & 0.089844 & -0.068359 & -0.015625 & -0.05518 & 0.001953 & -0.015625\\
$\overline{\Pi}_{3,1}$[24] & 0         & 0 & 0 & 0.010149 & -0.013532 & 0.00254 & -0.003383 & 0.010149\\
$\overline{\Pi}_{3,1}$[24] & 0         & 0.015625 & -0.058594 & 0.005859 & -0.015625 & 0.01025 & -0.005859 & 0.008789\\
$\overline{\Pi}_{3,1}$[8]  & 0         & 0 & 0 & -0.010149 & 0.013532 & 0.00761 & 0.02368 & 0.005074\\
$\overline{\Pi}_{3,2}$[12] & -0.015625 & 0.03125 & -0.015625 & 0.019531 & 0.015625 & -0.05078 & -0.033203 & -0.050781\\
$\overline{\Pi}_{3,2}$[24] & 0         & 0 & 0 & 0.010149 & -0.036084 & -0.01015 & 0 & -0.025372\\
$\overline{\Pi}_{3,2}$[12] & 0         & 0 & 0 & -0.040595 & -0.027063 & -0.02030 & -0.003383 & 0.030446\\
$\overline{\Pi}_{3,2}$[24] & 0         & -0.03125 & 0.046875 & 0.005859 & 0.007813 & 0 & 0.003906 & -0.008789\\
$\overline{\Pi}_{3,2}$[12] & 0         & -0.03125 & 0.078125 & 0.007813 & -0.036458 & -0.01172 & 0.013672 & -0.017578\\
$\overline{\Pi}_{3,2}$[12] & 0         & 0 & 0 & 0.006766 & 0 & 0 & 0.001128 & -0.020297\\
\end{tabular}
\end{ruledtabular}
\end{table*}

\section{First-principles methodology}\label{sec:fp_methodology}
The Vienna \emph{Ab initio} Simulation Package (VASP)\cite{1996Kre} was used to
perform the density functional theory (DFT) electronic structure calculations.
The projector augmented wave (PAW) method\cite{1999Kre} was chosen and the
generalized gradient approximation (GGA)\cite{1992Per} was used to take into
account exchange and correlation contributions to the Hamiltonian of the
ion-electron system. An energy cutoff of 364 $eV$ was used to calculate the
electronic structures of all the SQSs. 5,000 {\em k}-points per reciprocal atom
based on the Monkhorst-Pack scheme for the Brillouin-zone sampling was used.
In all first-principles calculations of binary and ternary SQSs in the present
work, structures are relaxed in two ways: full relaxation and volume relaxation
to preserve the fcc symmetry. It should be emphasized here that the preserved
symmetry is that of SQS when all the atoms are substituted into a single
element, which is the underlying fcc symmetry. For the full relaxation of SQS,
structures are completely relaxed with respect to all degrees of freedom i.e.
cell shape, volume, and ionic positions, while the symmetry-preserving
relaxation only allows to change the cell volume for cubic structures, such as
fcc and bcc. In calculating hcp SQSs, however, relaxing only the volume will
fix the $c/a$ ratio of the underlying hcp symmetry\cite{2006Shi}. Therefore,
the shape of SQSs has to be relaxed as well as the volume in hcp SQSs
calculations. Since the present work focuses on the calculations of fcc SQSs,
symmetry-preserving relaxation is equivalent to volume relaxation. For
symmetry-preserved calculations, all the atoms are still at the lattice sites
of fcc's and only the \emph{effective} lattice parameter of fcc changes with
this constrained relaxation scheme. However, the local relaxation due to the
like- and dislike-bondings cannot be taken into account. Further discussion
with regard to the different relaxation schemes of SQSs can be found in Ref.
\cite{2006Shi}.

\section{Results and discussions}\label{sec:results_discussions}
The mixing enthalpy derived from first-principles calculations of ternary fcc
SQSs should be compared to relevant experimental measurements. However, there
are always ordering effects at low temperatures where the fcc solid solutions
show complete solubility in binaries, such as Cu-Au and Au-Pd systems. Due to
such ordering in binaries, it is almost impossible to reliably determine the
mixing enthalpy for ternary fcc solid solutions from experiments.

In this work, the Ca-Sr-Yb system has been selected to apply the generated
ternary fcc SQSs which presumably has complete solubility in the fcc phase for
all binaries and ternary without any reported order/disorder transition. Both
the Ca-Sr and Ca-Yb systems show complete solubility for both fcc and bcc
phases at low and high temperatures respectively without the formation of any
intermetallic compounds\cite{1986Alc,1987Gsc}. There is no reported phase
diagram for the Sr-Yb system, however, from the similarity of the two binary
systems, Ca-Sr and Ca-Yb, it can be postulated that Sr-Yb also would have
complete solubility for both fcc and bcc phases. Consequently, it may
cautiously be expected that the combined ternary, the Ca-Sr-Yb system, would
have the fcc solid solution phase throughout the entire composition range at
low temperatures.

\subsection{Binary SQSs for the Ca-Sr-Yb system}
\begin{figure*}[htb]
\centering%
\mbox
{ %
    \subfigure[~Ca-Sr (liquid)]{%
        \label{fig:casrliq}
        \includegraphics[width=2.5in]{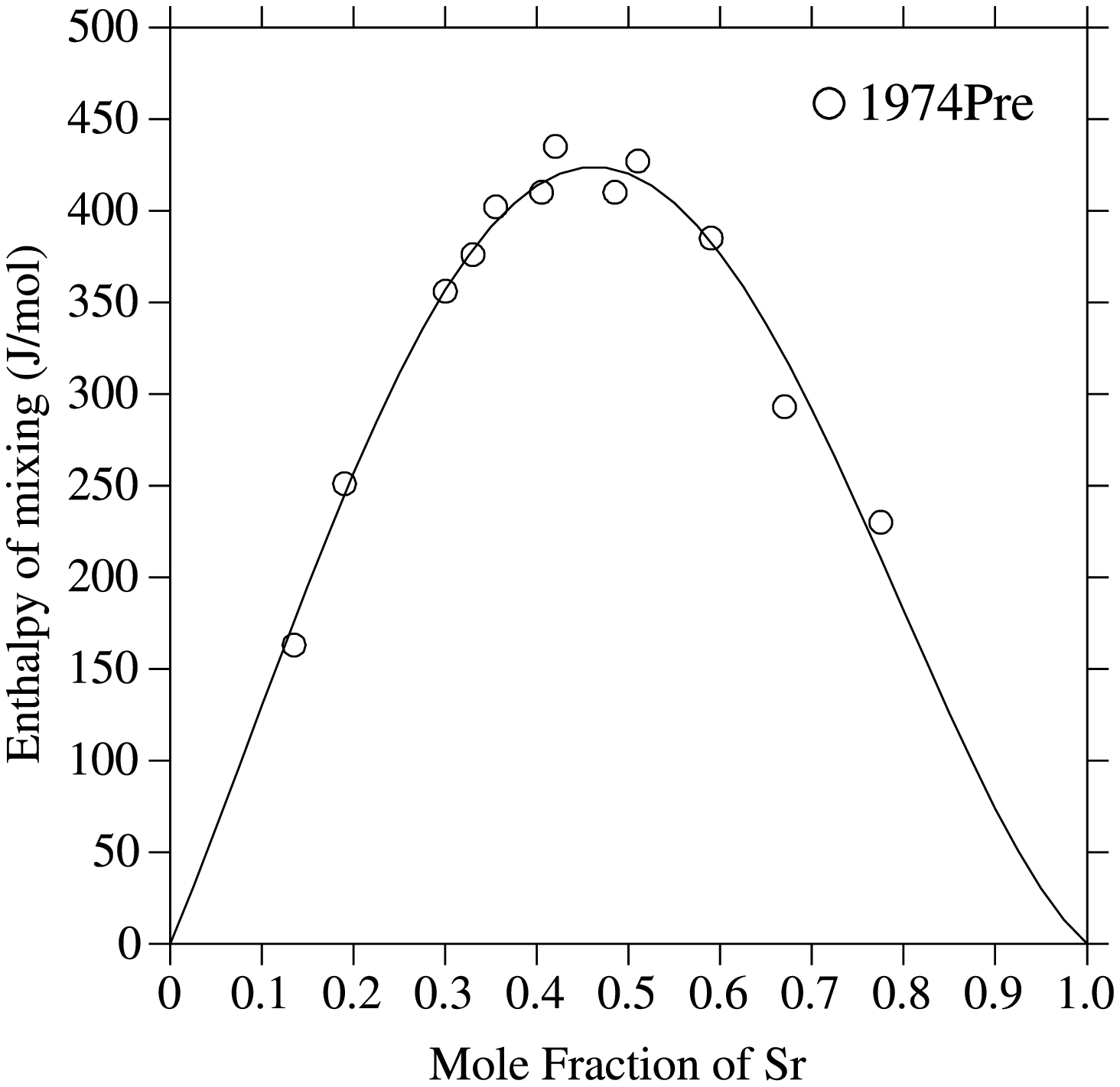}\quad
    }%
    \subfigure[~Ca-Sr (fcc)]{%
        \label{fig:casrfcc}
        \includegraphics[width=2.5in]{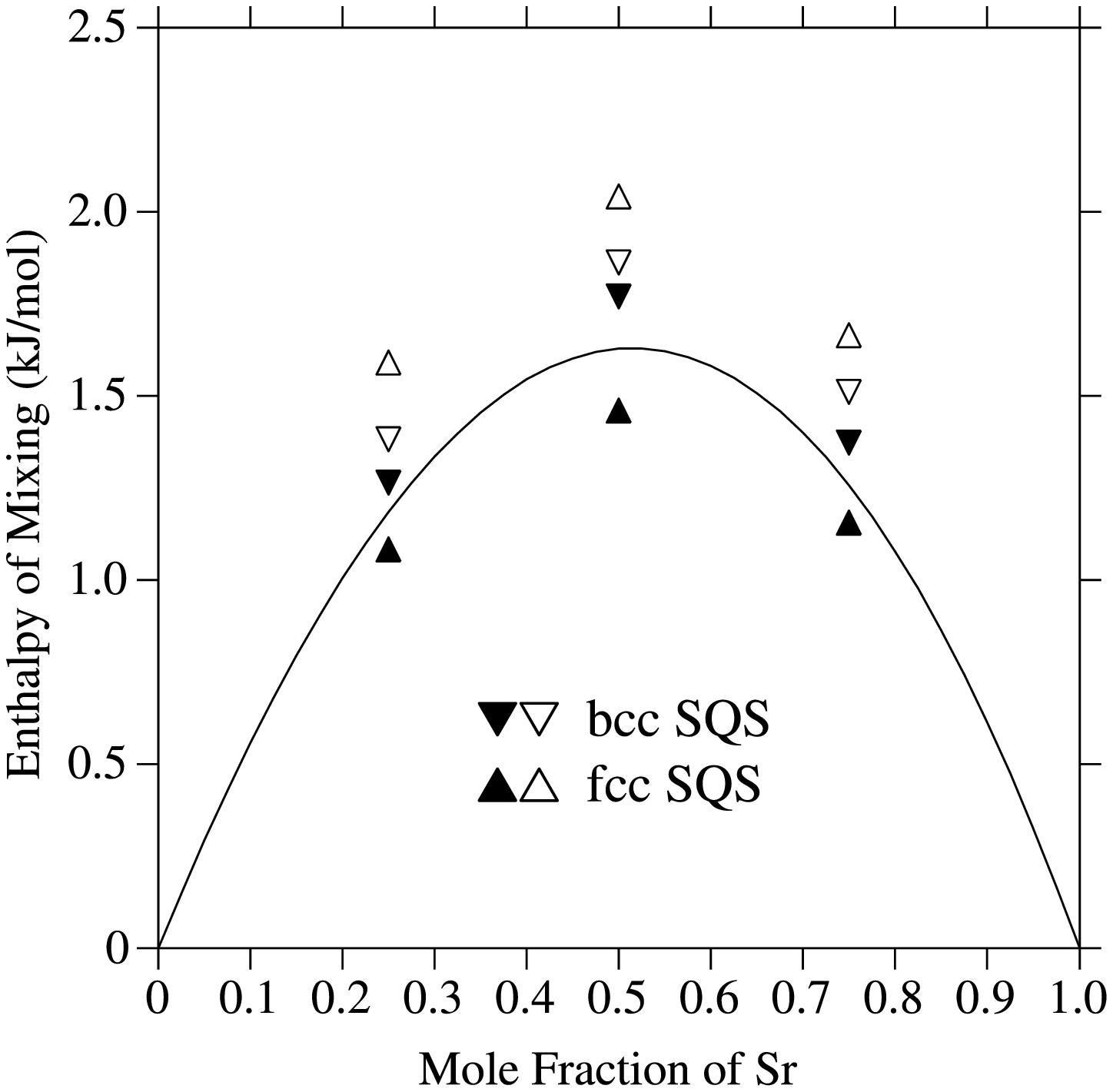}
    }
}\\%
\mbox
{ %
    \subfigure[~Sr-Yb (fcc)]{%
        \label{fig:srybfcc}
        \includegraphics[width=2.5in]{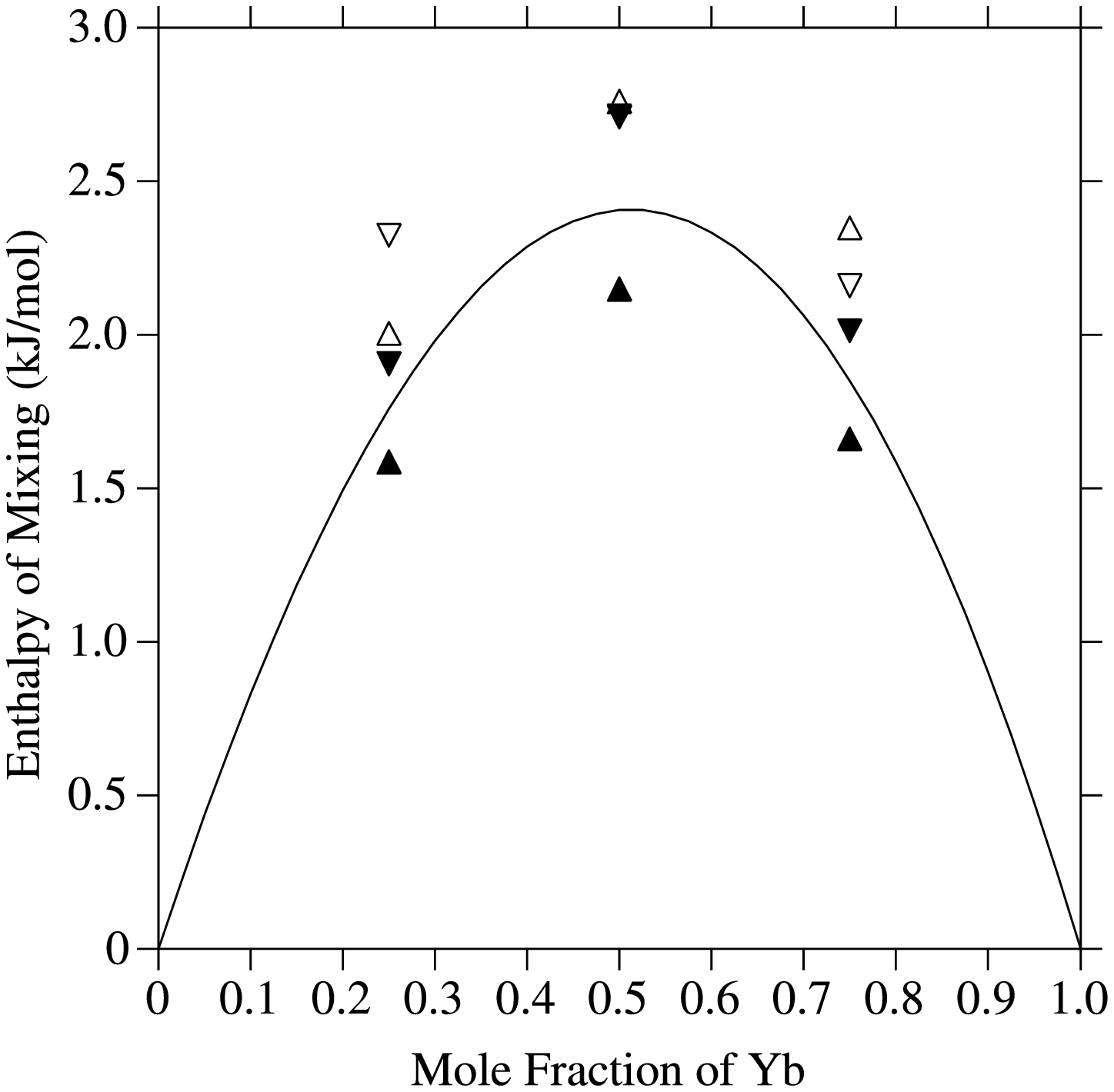}\quad
    }%
    \subfigure[~Yb-Ca (fcc)]{%
        \label{fig:ybcafcc}
        \includegraphics[width=2.5in]{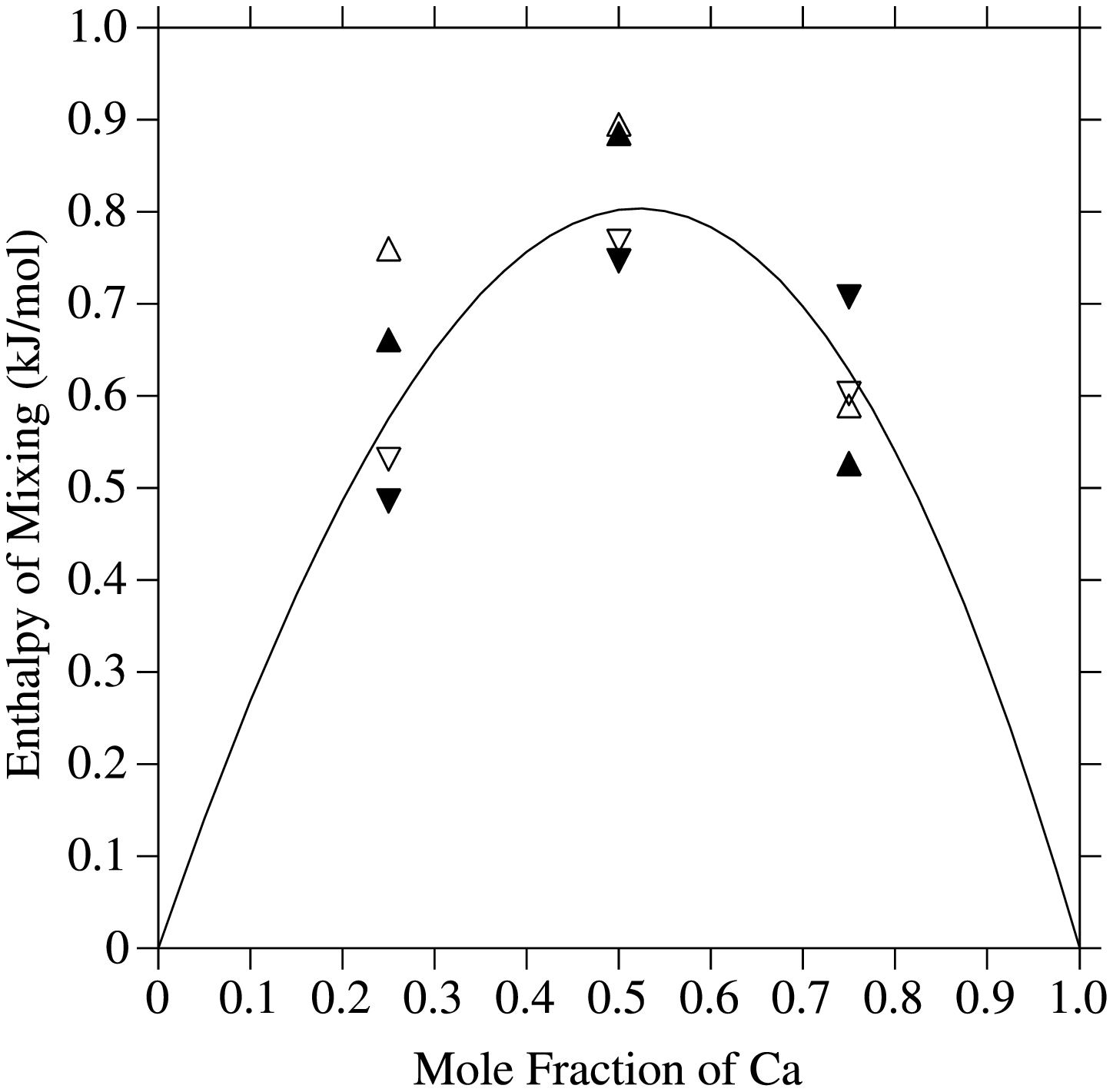}
    }
}%
\caption{Calculated mixing enthalpies for (a) the liquid phase in the Ca-Sr
with experimental measurement\cite{1974Pre} and the fcc phase for (b) Ca-Sr,
(c) Sr-Yb, and (d) Yb-Ca with first-principles calculations of fully relaxed
SQSs. Note that fcc and bcc phases are evaluated with same model parameters.
Open and closed symbols represent symmetry preserved and fully relaxed
calculations, respectively.}
\label{fig:casryb_bin} %
\end{figure*}

Prior to applying the ternary SQSs to the Ca-Sr-Yb system, the mixing behavior
of the fcc phase in its constitutive three binaries was investigated through
8-atom binary fcc SQSs at three different compositions, namely $x$=0.25, 0.5,
and 0.75 in $A_{1-x}B_x$ alloys. Calculated mixing enthalpies from binary fcc
SQSs are combined with experimental data from the
literature\cite{1958Sch,1974Pre} to evaluate interaction parameters in Eqn.
\ref{eqn:xs2_interaction} for each binary. For the sake of simplicity,
parameters for bcc have been modeled as identical to those of fcc. Binary bcc
SQSs also have been calculated and included in the parameter evaluation
process. FIG. \ref{fig:casryb_bin} shows that the formation energies of binary
bcc and fcc SQSs very close to each other in all cases. The congruent melting
of bcc is observed in both the Ca-Sr and Yb-Ca systems, thus the Sr-Yb system
has been modeled to have it as well on the assumption that Sr-Yb would behave
similarly.
Calculated mixing enthalpies for fcc and bcc phases of the three binaries are
shown in FIG. \ref{fig:casryb_bin} with first-principles calculations of binary
fcc and bcc SQSs. Mixing enthalpy of the liquid phase in Ca-Sr is also
calculated and compared with experimental measurement\cite{1974Pre}. Structural
analysis of binary SQS calculations shows that the local relaxation effect in
both fcc and bcc SQSs is small after the full relaxation and the difference
between fully relaxed and symmetry preserved structures are at most $\sim$1
kJ/mol.

%
It is intriguing to see that only the Yb-Ca system has the mixing enthalpy
close to zero among three binaries in FIG. \ref{fig:casryb_bin}, which implies
that Yb-Ca is likely to have ideal mixing in both fcc and bcc phases. The ideal
mixing of Yb and Ca is attributed to their similar lattice parameters. The
other two systems, Ca-Sr and Sr-Yb, have rather positive ($\sim$2 kJ/mol)
mixing enthalpies and significant lattice parameter differences ($a_{\rm
Ca}\sim a_{\rm Yb} \ll a_{\rm Sr}$), which can be the reason for the non-ideal
mixing behavior, unlike Yb-Ca.

\begin{figure}[htb]
\centering%
\includegraphics[width=3.0in]{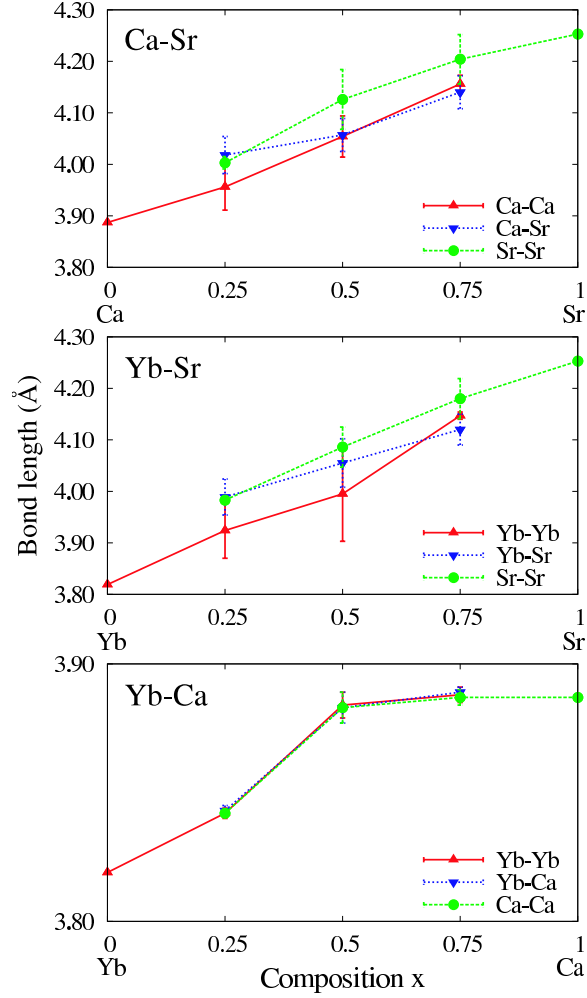}\\
\caption{\label{fig:bondlength_bin}%
SQS bond lengths for three binaries, Ca-Sr, Yb-Sr, and Yb-Ca. Error bars
correspond to the standard deviation of the bond length distributions.
}%
\end{figure}

The bond length analysis for the fully relaxed SQSs in FIG.
\ref{fig:bondlength_bin} shows that first nearest-neighbor average bond lengths
follow Vegard's law closely in all calculations. This observation indicates
that the lattice parameter of the fcc solid solution varies linearly with the
composition change and there is no significant geometrical distortion.

\subsection{Ternary fcc SQSs for the Ca-Sr-Yb system}
First-principles calculations of ternary fcc SQSs at four different
compositions in the Ca-Sr-Yb system, namely $x_{\rm Ca}=x_{\rm Sr}=x_{\rm
Yb}=\tfrac{1}{3}$; $x_{\rm Ca}=\tfrac{1}{2}$, $x_{\rm Sr}=x_{\rm
Yb}=\tfrac{1}{4}$; $x_{\rm Sr}=\tfrac{1}{2}$, $x_{\rm Ca}=x_{\rm
Yb}=\tfrac{1}{4}$; and $x_{\rm Yb}=\tfrac{1}{2}$, $x_{\rm Ca}=x_{\rm
Sr}=\tfrac{1}{4}$, have been considered to investigate the ternary
interactions. Three isoplethal sections, connecting the equimolar composition
and three other compositions when $x_i=1/2$, $x_j/x_k=1$, are selected to
examine the mixing enthalpy for the Ca-Sr-Yb ternary system.

\begin{figure}[htb]
\centering%
    \includegraphics[width=3.0in]{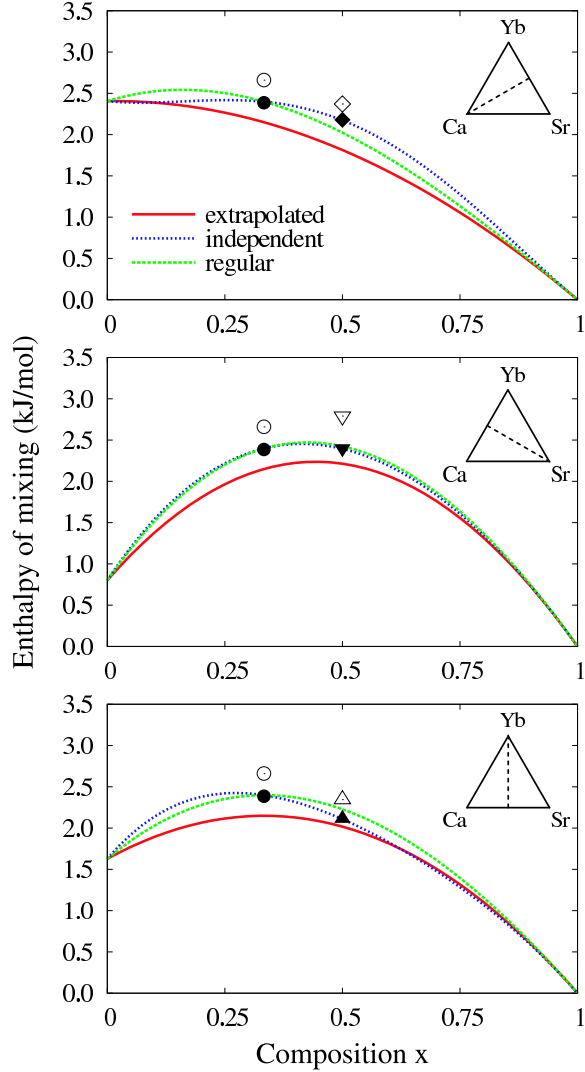}\\
\caption{\label{fig:casryb_tern}%
Calculated mixing enthalpies for the fcc phase in the Ca-Sr-Yb system with
first-principles calculations results of ternary SQSs in three pseudo-binaries,
Ca-SrYb, Sr-YbCa, and Yb-CaSr.
Extrapolated mixing enthalpy for the fcc phase from three binaries is
compared with the ones with ternary interaction parameters.
Open and closed symbols represent symmetry preserved and fully relaxed SQS
calculations, respectively.
}%
\end{figure}

\begin{figure*}[htbp]
\centering %
    \subfigure[~Radial distribution of $\rm Ca_1Sr_1Yb_1$]{%
        \includegraphics[height=2.25in]{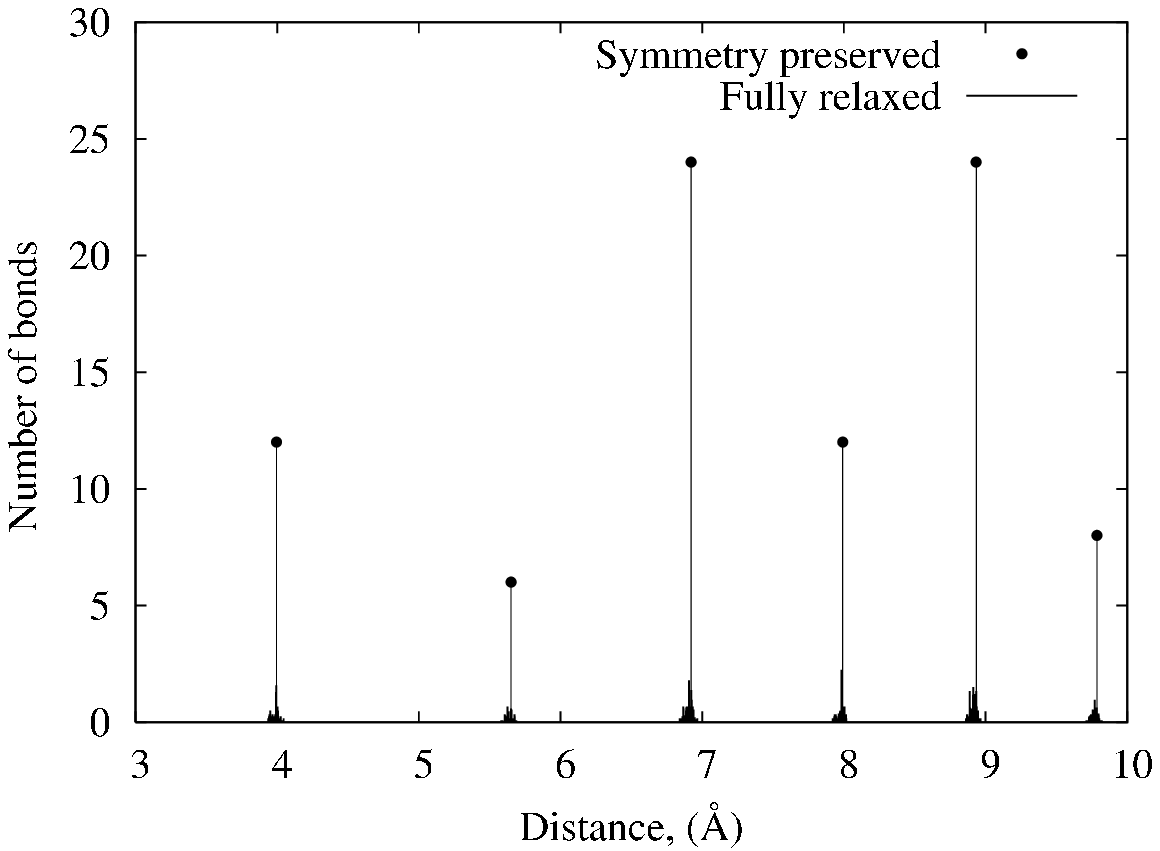} \quad
        \label{fig:rdf_casryb}
    }%
    \subfigure[~Smoothed and fitted RD's of fully relaxed $\rm Ca_1Sr_1Yb_1$]{%
        \includegraphics[height=2.25in]{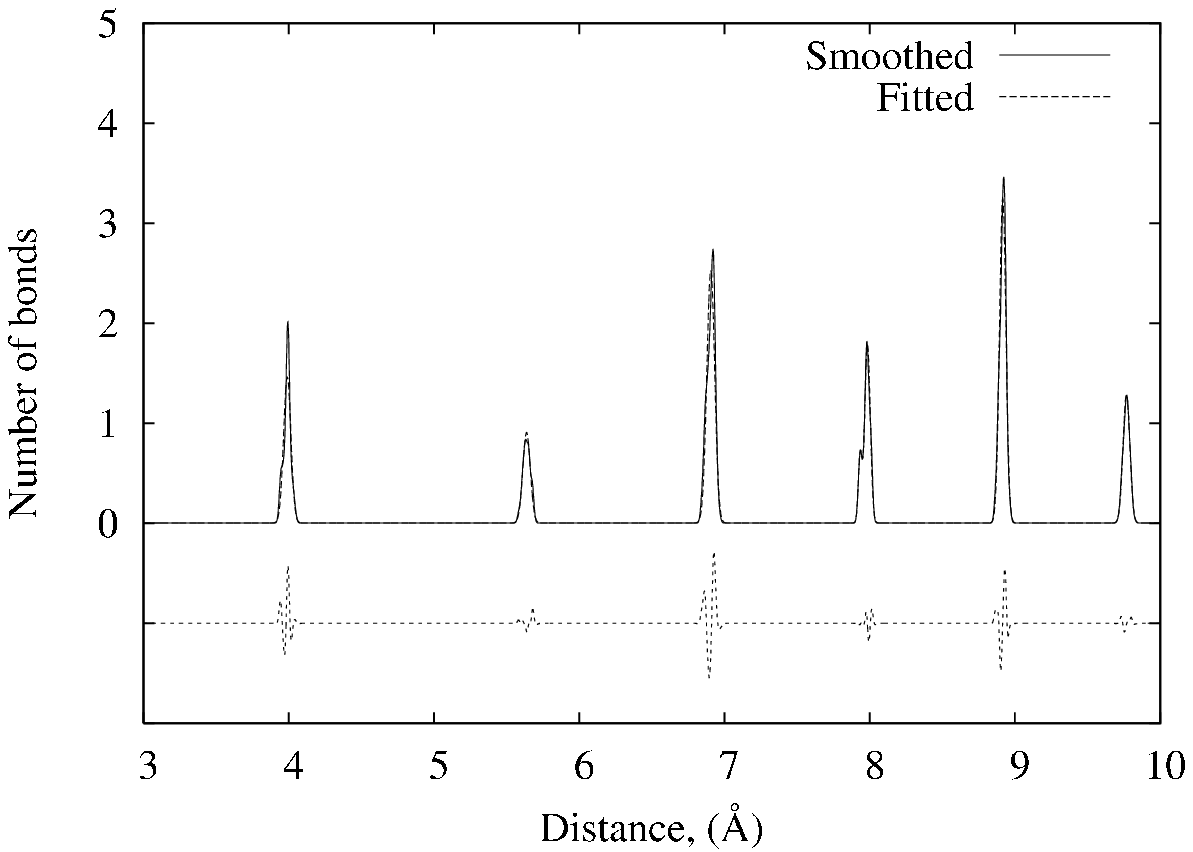}%
        \label{fig:rdcasrybfit}
    }\\%
\caption{\label{fig:rdfcasryb}%
Radial distribution analysis of $\rm Ca_1Sr_1Yb_1$ ternary fcc SQSs. The dotted
line in (b) under the smoothed and fitted curves are the error between the two
curves.}
\end{figure*}

Calculated enthalpies of mixing from ternary fcc SQSs are shown in FIG.
\ref{fig:casryb_tern}, including extrapolated results from the three binaries
and improved enthalpies of mixing to reproduce ternary fcc SQSs results by
introducing ternary interaction parameters. All the fully relaxed ternary SQSs
show that the effect of local relaxation is also small as in its constitutive
binaries, thus the energy differences between the symmetry preserved and fully
relaxed calculations are quite small. FIG. \ref{fig:rdfcasryb} shows the radial
distribution analysis of the fully relaxed SQS at the equimolar composition.
The narrow distribution along each of the bond-lengths indicates that the
effect of local relaxation is small. As shown in FIG. \ref{fig:casryb_tern},
energy differences between fully relaxed and symmetry preserved calculations
are small($\sim$ 0.5 kJ/mol).

As can be seen in FIG. \ref{fig:casryb_tern}, the mixing enthalpy extrapolated
from the three binaries is slightly lower than that derived from
first-principles calculations of ternary fcc SQSs. Thus, ternary interaction
parameters are introduced to improve the ternary mixing enthalpy. According to
Eqn. \ref{eqn:xs2_tern}, the contribution from the ternary excess Gibbs energy
for the fcc phase in the Ca-Sr-Yb system can be denoted as

\begin{equation}\label{eqn:tern_casryb}
^{xs}G^{tern,{\rm fcc}} = x_{\rm Ca}x_{\rm Sr}x_{\rm Yb} ( L^{\rm fcc}_{\rm
Ca}x_{\rm Ca}+L^{\rm fcc}_{\rm Sr}x_{\rm Sr}+L^{\rm fcc}_{\rm Yb}x_{\rm Yb})
\end{equation}

\noindent or

\begin{equation}\label{eqn:simple_casryb}
^{xs}G^{tern,{\rm fcc}}=x_{\rm Ca }x_{\rm Sr }x_{\rm Yb}L^{\rm fcc}_{\rm
CaSrYb}
\end{equation}

\noindent as simplified in Eqn. \ref{eqn:xs_simple} when the ternary fcc is
considered as a regular solution. When three independent ternary interaction
parameters ($L_{\rm Ca}=25940$, $L_{\rm Sr}=2913$, and $L_{\rm Yb}=-8645$
J/mol) are used, slightly better agreement with ternary SQSs was made than with
a single regular interaction parameter ($L_{\rm CaSrYb}=6736$ J/mol). The
calculated mixing enthalpy at the equimolar composition are evaluated as the
same value regardless of the interaction parameters since all the data are
equally weighted.

\begin{figure*}[htbp]
\centering%
    \includegraphics[width=4.5in]{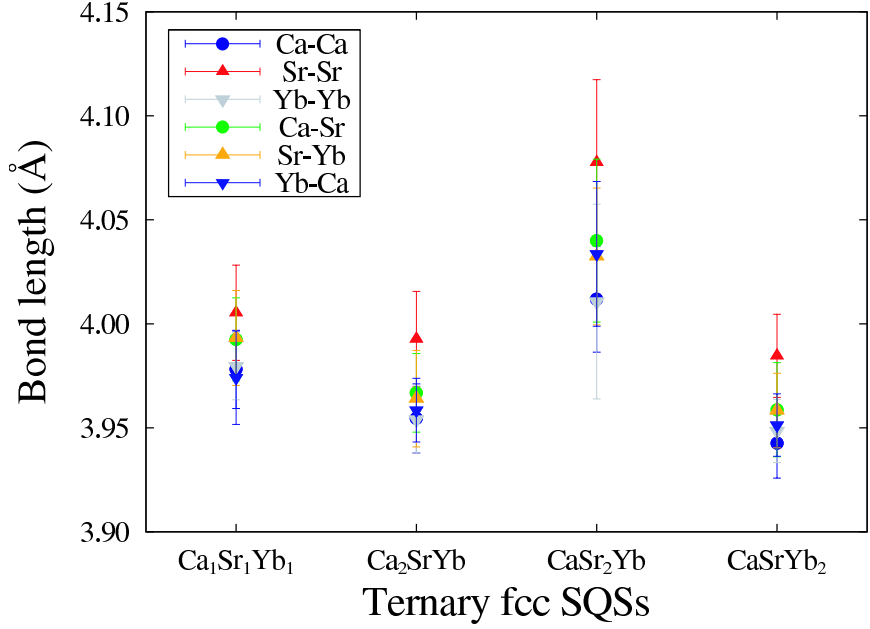}
\caption{%
First nearest-neighbor average bond lengths for the fully relaxed ternary fcc
SQSs in the Ca-Sr-Yb system. Error bar corresponds to the standard deviation
of the bond length distributions.
}%
\label{fig:tern_bond}%
\end{figure*}

Six different bond lengths, three like-bondings and three dislike-bondings, of
four fully relaxed SQSs have been analyzed. In FIG. \ref{fig:tern_bond} the
bond lengths corresponding to the first nearest neighbors for all SQSs are
presented. In all SQS calculations, the Sr-Sr bonding is always the longest and
this is attributed to the biggest lattice parameter of Sr among the three
elements.
%

\begin{figure*}[htbp]
\centering%
    \subfigure[~Ca-SrYb]{%
        \label{fig:caDOS}
        \includegraphics[width=2.0in]{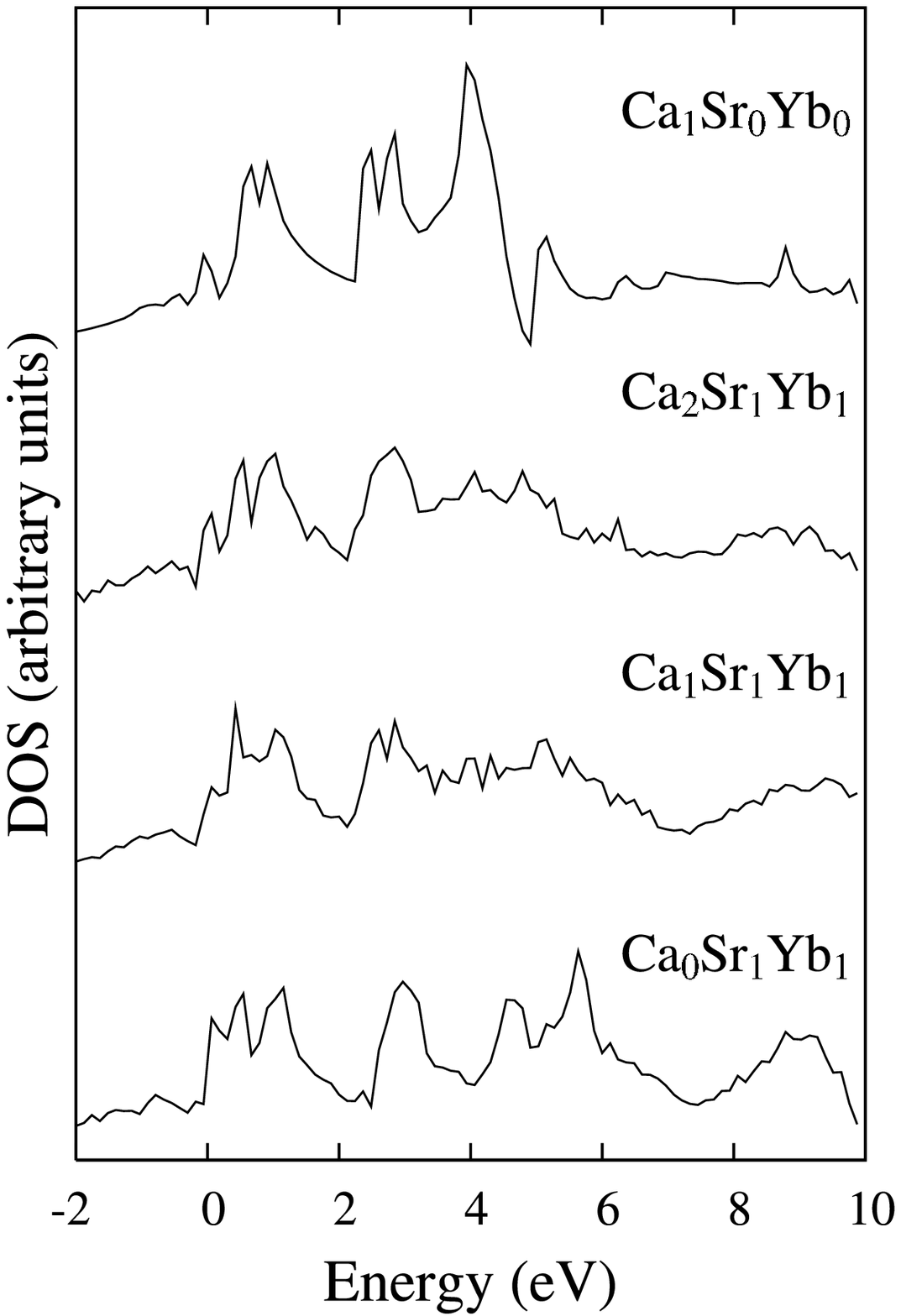}
    }
    \subfigure[~Sr-YbCa]{%
        \label{fig:srDOS}
        \includegraphics[width=2.0in]{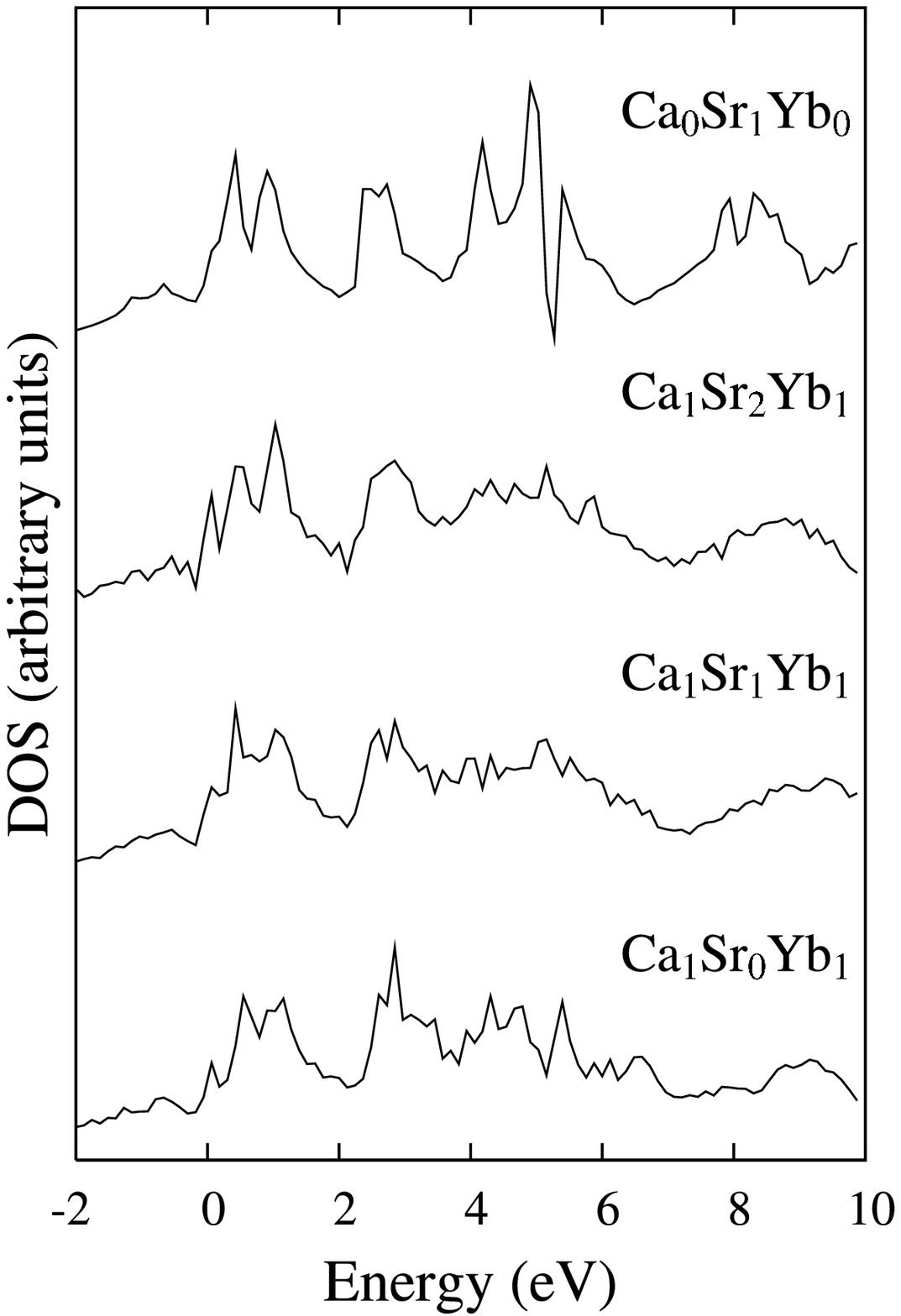}
    }
    \subfigure[~Yb-CaSr]{%
        \label{fig:ybDOS}
        \includegraphics[width=2.0in]{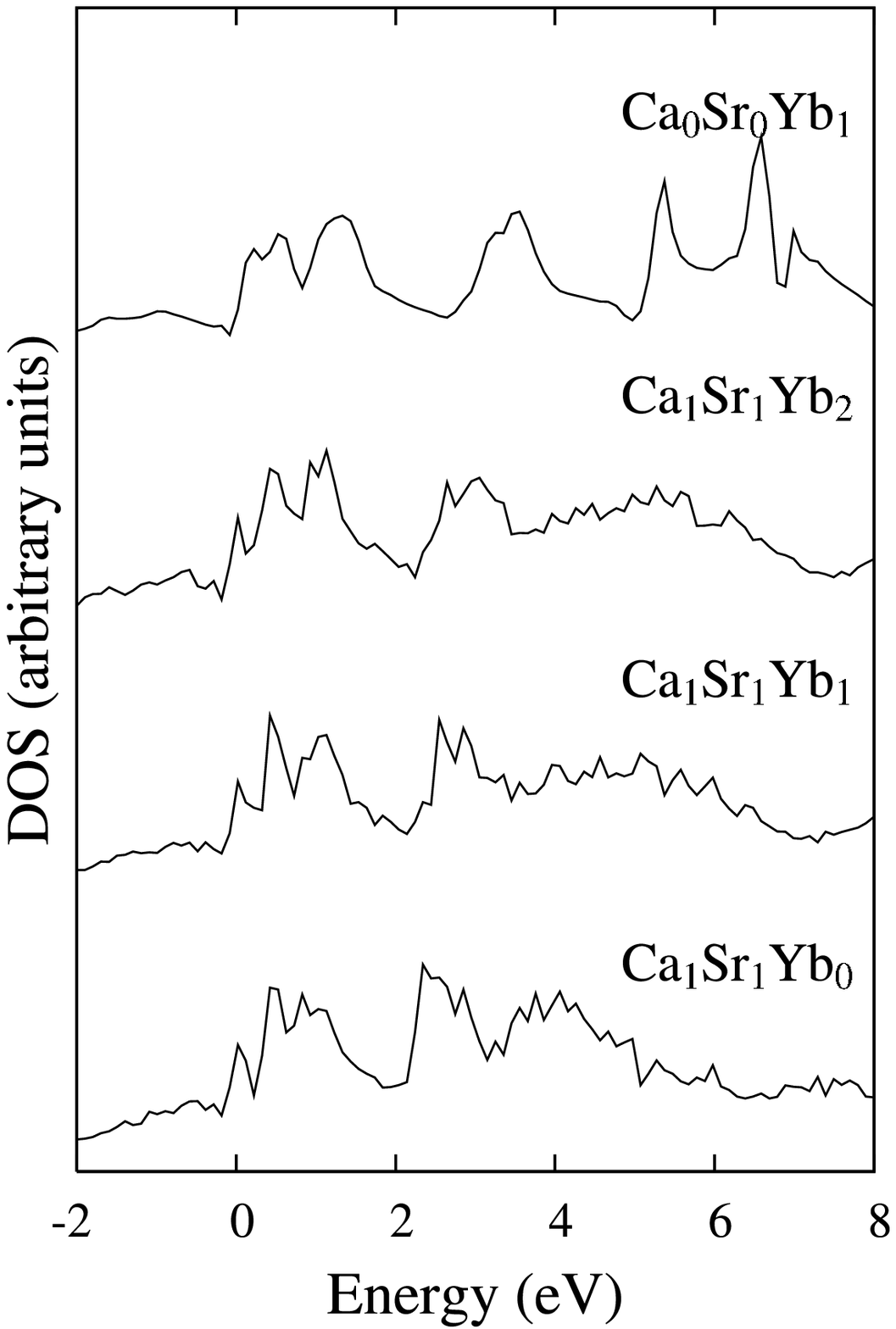}
    }
\caption{\label{fig:DOS}%
Calculated density of states for Ca-Sr-Yb as three pseudo-binaries.
}%
\end{figure*}

As a final analysis of the ability of the generated SQSs to reproduce the
properties of ternary fcc solid solutions, FIG. \ref{fig:DOS} shows the
alloying effects on the electronic DOS in the Ca-Sr-Yb system. Three
pseudo-binaries, connecting a pure element and equimolar composition of the
other two elements, have been selected. Since all three elements have the same
number of valence electrons, significant changes in the electronic DOS were not
observed. Instead, the electronic DOS of the outer states gradually transformed
into that of the pure element as the composition changed towards the pure
element. The peak around $\sim$ 9 $eV$ of SrYb in FIG. \ref{fig:caDOS} flattens
out as Ca increases, and the peaks around $\sim$ 8 $eV$ and $\sim$ 7 $eV$ are
become more pronounced as the content of Sr and Yb increases as shown in FIGs.
\ref{fig:srDOS} and \ref{fig:ybDOS}, respectively.

\section{Conclusion}\label{sec:conclusion}
In the present work, two ternary fcc SQSs at different compositions,
$x_A=x_B=x_C=\tfrac{1}{3}$ and $x_A=\tfrac{1}{2}$, $x_B=x_C=\tfrac{1}{4}$, are
generated and their correlation functions are satisfactorily close to those of
random fcc solid solutions. The generated SQSs are applied to the Ca-Sr-Yb
system which presumably has a complete solubility range without order/disorder
transitions in ternary fcc solid solutions. Mixing enthalpies for the fcc phase
in three binaries are evaluated from first-principles calculations of fcc and
bcc SQSs and available experimental data with the CALPHAD approach. It is found
that the local relaxation effect of fcc and bcc phases are very small and
mixing enthalpies are slightly positive in all cases. Evaluated mixing
enthalpies for the fcc phase in three binaries are then extrapolated to the
ternary system.

First-principles results of four ternary SQSs at $x_{\rm Ca}=x_{\rm Sr}=x_{\rm
Yb}=\tfrac{1}{3}$; $x_{\rm Ca}=\tfrac{1}{2}$, $x_{\rm Sr}=x_{\rm
Yb}=\tfrac{1}{4}$; $x_{\rm Sr}=\tfrac{1}{2}$, $x_{\rm Ca}=x_{\rm
Yb}=\tfrac{1}{4}$; and $x_{\rm Yb}=\tfrac{1}{2}$, $x_{\rm Ca}=x_{\rm
Sr}=\tfrac{1}{4}$ show that the local relaxation effect is also very small in
the ternary system. Extrapolated mixing enthalpy from its constitutive binaries
are slightly lower than those from first-principles calculations of ternary fcc
SQSs. Thus ternary interaction parameters for fcc solid solution phases are
introduced to further improve the ternary mixing enthalpy. It can be concluded
that the generated ternary fcc SQSs are able to reproduce thermodynamic
properties of ternary fcc solid solutions and can readily be applied to other
systems.

\begin{acknowledgements}
This work is funded by the National Science Foundation (NSF) through Grant No.
DMR-0205232. First-principles calculations were carried out on the LION
clusters at the Pennsylvania State University supported in part by the NSF
grants (DMR-9983532, DMR-0122638, and DMR-0205232) and in part by the Materials
Simulation Center and the Graduate Education and Research Services at the
Pennsylvania State University.
\end{acknowledgements}%

\appendix%

\section{Evaluated interaction parameters}
Interaction parameters for the liquid and solid solution phases (fcc and bcc)
in the three binaries of the Ca-Sr-Yb system ---as evaluated from the CALPHAD
modeling in the present work--- are listed in TABLE \ref{tbl:parameters}. 
Notations are explained in Eqn. \ref{eqn:xs2_bin}.

\begin{table}[h]
\fontsize{9}{9pt}\selectfont 
\caption{\label{tbl:parameters}%
Evaluated binary interaction parameters for the Ca-Sr-Yb system (all in S.I.
units). Gibbs energies for pure elements are from the SGTE pure element
database\cite{1991Din}.}
\begin{ruledtabular}
\begin{tabular}{ccl}
Systems & Phases & Evaluated parameters\\
\hline %
Ca-Sr   & Liquid  & $^0L=1680+2.68T$, $^1L=388-1.153T$, $^2L=-856+.631T$ \\
        & fcc,bcc & $^0L=6511$, $^1L=-382$ \\
\hline%
Sr-Yb   & Liquid  & $^0L=8850$ \\
        & fcc,bcc & $^0L=9624$, $^1L=-488$ \\
\hline%
Yb-Ca   & Liquid  & $^0L=2689$, $^1L=676$\\
        & fcc,bcc & $^0L=3207$, $^1L=280$
\end{tabular}
\end{ruledtabular}
\end{table}

\section{Special quasirandom structures for the ternary fcc solution phase}
Special quasirandom structures are $N$-atom per cell periodic structures
designed to have correlation functions close to those of completely random
alloys. The ternary fcc SQSs used in this work are presented. Lattice vectors
are given as $\textbf{a}$, $\textbf{b}$, and $\textbf{c}$, and atom positions
of $A$, $B$, and $C$ atoms are given as $A_i$, $B_i$, and $C_i$, respectively.

\begin{table}[htbp]
\centering%
\fontsize{9}{9pt}\selectfont 
\caption{\label{tbl:lattice}%
Structural descriptions of the SQSs for the ternary fcc solid solution. Lattice
vectors and atomic positions are given in fractional coordinates of the fcc
lattice. Atomic positions are given for the ideal, unrelaxed fcc sites.
}%
\begin{ruledtabular}
\begin{tabular}{ccc}
$A_1B_1C_1$ (SQS-24) & \multicolumn{2}{c}{$A_2B_1C_1$ (SQS-32)}\\
Lattice vector & \multicolumn{2}{c}{Lattice vector}\\
$\left(\begin{array}{rrrr}
   3 & 1 & -1 \\
  -3 & 1 & -1 \\
   0 & \frac{1}{2} & \frac{1}{2} \\
\end{array}\right)$
&
\multicolumn{2}{c}{$\left(\begin{array}{rrrr}
   1 & 1 &  2 \\
   1 & 1 & -2 \\
  -1 & 1 &  0 \\
\end{array}\right)$}\\
\hline %
Atom positions & \multicolumn{2}{c}{Atom positions}\\
$\begin{array}[t]{rrrr}
1             & 1\frac{1}{2} & -\frac{1}{2} & A\\
\frac{1}{2}   & 2            & -1\frac{1}{2}  & A\\
0             & 1\frac{1}{2} & -\frac{1}{2}  & A\\
\frac{1}{2}   & \frac{1}{2}  & 0  & A\\
-1            & 2            & -1  & A\\
-1            & 1\frac{1}{2} & -\frac{1}{2}  & A\\
-\frac{1}{2}  & 1            & -\frac{1}{2}  & A\\
-1\frac{1}{2} & 1            & -\frac{1}{2}  & A\\
\end{array}  $
&
$\begin{array}[t]{rrrr}
 1\frac{1}{2} & 2            & \frac{1}{2}& A,\\
 0            & 2            & -1& A,\\
  \frac{1}{2} & 1\frac{1}{2} & -1& A,\\
 1            & 1\frac{1}{2} & 1\frac{1}{2}& A,\\
  \frac{1}{2} & 1            & -\frac{1}{2}& A,\\
 1            & 1\frac{1}{2} & -\frac{1}{2}& A,\\
 1            & 3            & 0& A,\\
 -\frac{1}{2} & 1\frac{1}{2} & 0& A,\\
\end{array}$
&
$\begin{array}[t]{rrrr}
 0            & 2            & 0  & A\\
 0            & 1\frac{1}{2} & -\frac{1}{2}  & A\\
  \frac{1}{2} & 2            & -\frac{1}{2}  & A\\
  \frac{1}{2} & 2\frac{1}{2} & -1  & A\\
 0            & 2            &  1  & A\\
  \frac{1}{2} & 2            & -1\frac{1}{2}  & A\\
  \frac{1}{2} & 2            & \frac{1}{2}  & A\\
 1            & 2            & -1  & A\\
\end{array}$
\\
\hline %
$\begin{array}[t]{rrrr}
 1            & 2            & -1  & B\\
 1\frac{1}{2} & 1\frac{1}{2} & -1  & B\\
 1\frac{1}{2} & 1            & -\frac{1}{2}  & B\\
 0            & 2            & -1  & B\\
  \frac{1}{2} & 1            & -\frac{1}{2}  & B\\
 -\frac{1}{2} & 2            & -1\frac{1}{2}  & B\\
 -1           & 1            & 0  & B\\
 -1\frac{1}{2}& 1\frac{1}{2} & -1  & B\\
\end{array}$
&
\multicolumn{2}{c}{
$\begin{array}[t]{rrrr}
  \frac{1}{2} & 2\frac{1}{2} & 1  & B\\
 1            & 2\frac{1}{2} &  \frac{1}{2}  & B\\
  \frac{1}{2} & 2            & 1\frac{1}{2}  & B\\
 1\frac{1}{2} & 2\frac{1}{2} & 0  & B\\
 1            & 2            & 0  & B\\
 1            & 1\frac{1}{2} & -1\frac{1}{2}  & B\\
  \frac{1}{2} & 1\frac{1}{2} & 1  & B\\
 1            & 2\frac{1}{2} & -\frac{1}{2}  & B\\
\end{array}$}
\\
\hline %
$\begin{array}[t]{rrrr}
 2            &  1\frac{1}{2} & -\frac{1}{2}  & C\\
 1            &  1            &  0  & C\\
  \frac{1}{2} &  1\frac{1}{2} & -1  & C\\
 0            &  1            &  0  & C\\
 0            &  2\frac{1}{2} & -1\frac{1}{2}  & C\\
 -\frac{1}{2} &  1\frac{1}{2} & -1  & C\\
 -\frac{1}{2} &  \frac{1}{2}  &  0  & C\\
 -2           &  1\frac{1}{2} &  -\frac{1}{2}  & C\\
\end{array}$
&
\multicolumn{2}{c}{
$\begin{array}[t]{rrrr}
 1            &  2            &  1  & C\\
  \frac{1}{2} &  2\frac{1}{2} &  0  & C\\
 0            &  1            &  0  & C\\
  \frac{1}{2} &  1\frac{1}{2} &  0  & C\\
  \frac{1}{2} &  1            &  \frac{1}{2}  & C\\
 1            &  1\frac{1}{2} &  \frac{1}{2}  & C\\
 0            &  1\frac{1}{2} &  \frac{1}{2}  & C\\
 1\frac{1}{2} &  2            &  -\frac{1}{2}  & C\\
\end{array}$}
\end{tabular}
\end{ruledtabular}
\end{table}

\end{document}